\documentclass[aps,prb,twocolumn,groupedaddress,showpacs,floatfix,superscriptaddress,longbibliography]{revtex4-2}
\usepackage[plainpages=false,pdfpagelabels,colorlinks=true,linkcolor=red,urlcolor=blue,citecolor=blue,pdftitle={Title},pdfauthor={},pdfdisplaydoctitle=true,pdfduplex=DuplexFlipLongEdge]{hyperref}
\usepackage{epsfig}
\usepackage{amsmath,amssymb}
\usepackage{physics}
\usepackage{graphicx}
\usepackage[dvipsnames,usenames]{color}
\usepackage[normalem]{ulem}
\usepackage{soul} % to use \st for strikeout

\renewcommand{\vec}[1]{\boldsymbol{#1}}

\begin{document}
\title{Charge excitations across a superconductor-insulator transition}

\author{Xiaodong Jin}
\affiliation{Beijing Computational Science Research Center, Beijing 100084, China}

\author{Yuhai Liu}
\affiliation{Beijing Computational Science Research Center, Beijing 100084, China}

\author{Rubem Mondaini}
\email{rmondaini@csrc.ac.cn}
\affiliation{Beijing Computational Science Research Center, Beijing 100084, China}

\author{Marcos Rigol}
\email{mrigol@phys.psu.edu}
\affiliation{Department of Physics, The Pennsylvania State University, University Park, PA 16802, USA}

\begin{abstract}
We study the superconductor-insulator transition (SIT) in the ground state of the attractive honeycomb Hubbard model in the presence of a staggered potential (a mass term), using a combination of unbiased computational methods, namely, exact diagonalization and quantum Monte Carlo simulations. We probe the nature of the lowest-energy charge excitations across the SIT and show that they are bosonic, as inferred (and shown in the strongly interacting regime) in a previous study of the same model in the square lattice. Increasing the strength of the staggered potential leads to a crossover in which bosonic low-energy excitations give way to fermionic ones within the insulating phase. We also show that the SIT belongs to the 3$d$-XY universality class, like in its square lattice counterpart. The robustness of our results in these two lattice geometries supports the expectation that our findings are universal for SITs in clean systems.
\end{abstract}

\maketitle

\section{Introduction}

The theory of conventional superconductors describes the emergence of pairing among repulsive charges by means of a Fermi surface instability in a metal (or Fermi liquid)~\cite{BCS1957}. It leads to an effective attractive interaction whose hallmark is the formation of a gap for one-particle excitations, while zero-momentum pair excitations are gapless~\cite{Tinkham2004}. Yet, various cases are known to evade this paradigm, where the parent state does not have a well-defined Fermi surface, such as in the case of insulators. This superconductor-insulator transition (SIT) has been experimentally investigated in various low-dimensional systems, either via disorder tuning~\cite{Shahar1992, Sacepe2011, Sherman2015,Roy2018}, introducing magnetic fields~\cite{Hebard1990, Yazdani1995, Zhang2021}, or introducing changes in the carrier density~\cite{Parendo2005, Caviglia2008, Bollinger2011}. The development of analog quantum simulators, comprising ultracold atoms trapped in optical lattices~\cite{Bloch2008, Esslinger2010, Schafer2020}, has opened a door to controllably study such SIT transitions in exquisitely clean strongly correlated systems. 

Motivated by understanding the nature of SITs in clean strongly correlated systems, as well as their possible experimental exploration in experiments with ultracold fermionic atoms, in a previous work~\cite{Mondaini2015} two of us (R.M. and M.R., in collaboration with P. Nikoli\'c) studied the SIT in the attractive Hubbard model in the presence of a staggered potential in the square lattice geometry. We showed that in the insulating phase in the strongly interacting regime, the lowest-energy charge excitations can be bosonic or fermionic, depending on the parameters chosen. We also showed that, for all the Hamiltonian parameters that could be studied using quantum Monte Carlo simulations, the SIT belongs to 3$d$-XY universality class. We concluded from those results that the lowest-energy charge excitations are bosonic in {\it both} the superconducting and insulating phases across the SIT, as argued using field-theory arguments in the weak-coupling limit~\cite{Nikolic2011, Nikolic2011b}. Such a bosonic-like insulator is akin to a pseudogap phase, where pre-formation of pairs occurs but such pairs are not phase correlated to exhibit superconductivity~\cite{Mondaini2015}. Lowest-energy bosonic charge excitations across an SIT were later argued to also occur in a model of two coupled triangular lattices~\cite{Loh2016, Hazra2020}. 

Here we study the SIT in the ground state of the attractive Hubbard model in the presence of a staggered potential in the honeycomb lattice geometry. Our main goal is to directly probe the nature of the lowest-energy charge excitations across the SIT, which we find to be bosonic, and study their crossover to being fermionic upon increasing the strength of the staggered potential in the insulating phase. For this, we compute the one-particle (fermionic) and two-particle (bosonic) gaps, which are more precise probes of the nature of the lowest energy excitations than the probes used in Ref.~\cite{Mondaini2015}. We also study the universality class of the SIT, which we find to be 3$d$-XY as in the square lattice case~\cite{Mondaini2015}. Our second goal, which together with potential experimental realizations motivated us to study the honeycomb lattice geometry, is to show that our conclusions are robust independently of the lattice geometry (Ref.~\cite{Mondaini2015} focused on the square-lattice geometry); and likely to be universal for SITs in clean systems. The fermionic model considered in this paper, except for the next-nearest-neighbor terms, was studied experimentally with ultracold atoms~\cite{Messer2015}. To properly account for the effects of quantum fluctuations and strong correlations in our model, we use two unbiased computational techniques, exact diagonalization and quantum Monte Carlo simulations. 

The presentation is organized as follows. In Sec.~\ref{sec:model}, we introduce the attractive Hubbard model in the presence of a staggered potential in the honeycomb lattice geometry, present its phase diagram, and discuss limiting regimes that can be solved analytically. We also briefly introduce the unbiased computational techniques used to study this model. In Sec.~\ref{sec:phasediagram}, we discuss how the phase diagram of the model is obtained using exact diagonalization and quantum Monte Carlo simulations. This is where we show that the SIT belongs to 3$d$-XY universality class. Section~\ref{sec:excitations} is devoted to studying the one-particle (fermionic) and two-particle (bosonic) charge excitations across the SIT and how the nature of the lowest-energy one changes from bosonic to fermionic upon increasing the strength of the staggered potential in the insulating phase. Our results are summarized in Sec.~\ref{sec:summary}.

\section{Model and Methods}\label{sec:model}

\begin{figure}[!tb] %%% FIG1
    \includegraphics[width=0.98\columnwidth]{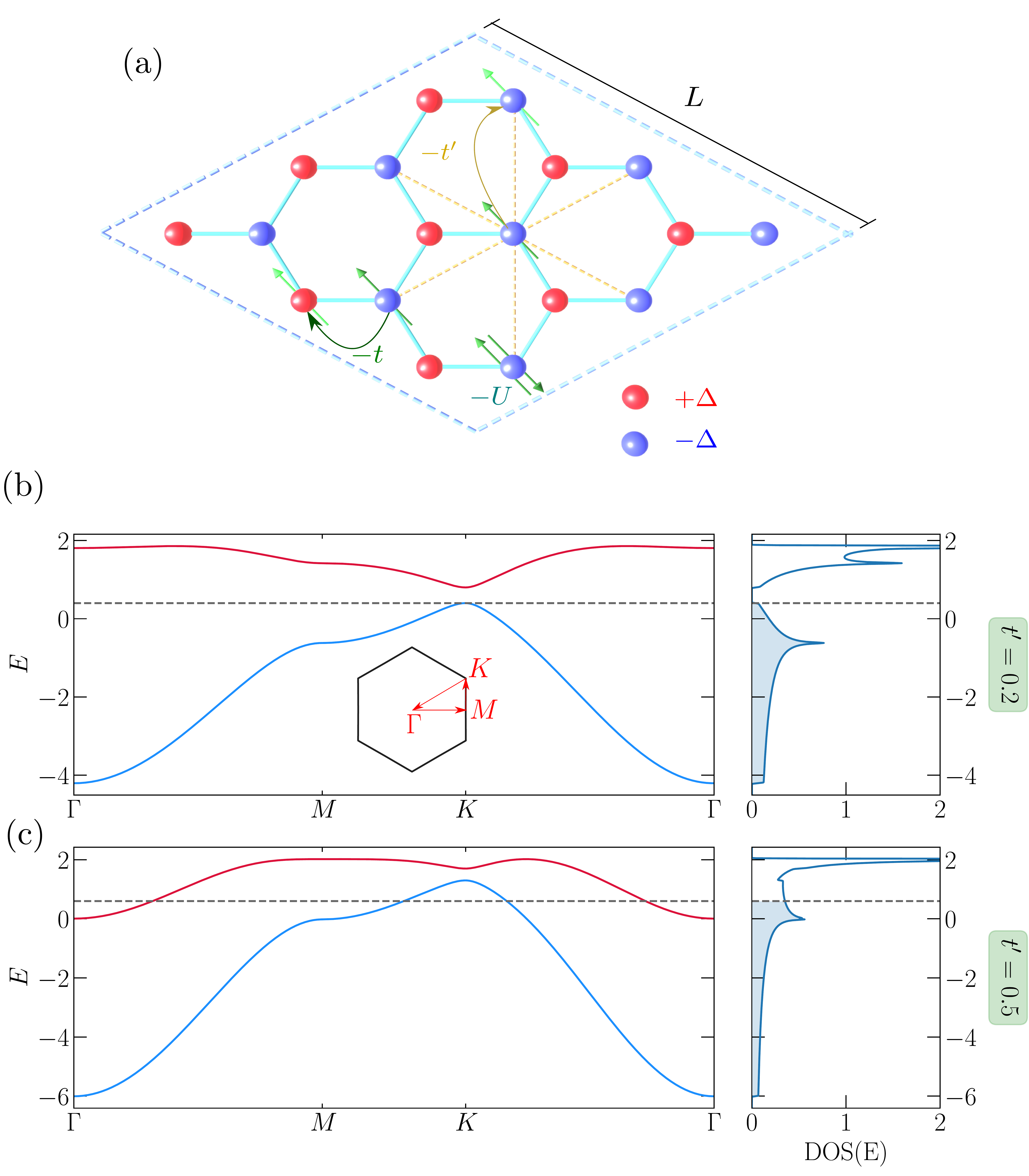}
    \caption{(a) Schematic representation of the terms in Eq.~\eqref{eq:Hamiltonian} in a honeycomb lattice with $V=2L^2$ sites for $L=3$. (b), (c) Band structure in the noninteracting ($U=0$) limit (left), and the corresponding density of states (right), for a strength of the staggered potential $\Delta = 0.2$. (b) Results for $t^\prime = 0.2$, for which the ground state is a band insulator. (c) Results for $t^\prime = 0.5$, for which the ground state is a metal. $t=1$ sets the energy scale.}
  \label{fig:cartoon_bands_dos}
\end{figure}

Our model of interest is the SU(2) honeycomb Hubbard model in the presence of a staggered potential~\cite{Messer2015},
\begin{eqnarray}
\label{eq:Hamiltonian}
    \hat H &=& -t \sum_{\langle {\vec {i\,j}} \rangle ,\sigma} \hat c^{\dagger}_{\vec{i}\sigma} \hat c^{}_{\vec{j}\sigma} -t^\prime \sum_{\langle\langle {\vec {i\,j}} \rangle\rangle ,\sigma} \hat c_{\vec{i}\sigma}^\dagger \hat c_{\vec{j}\sigma}^{} \\
    && + U \sum_{\vec i} \left(\hat n_{\vec i\uparrow}-\frac{1}{2}\right) \left(\hat n_{\vec i \downarrow}-\frac{1}{2}\right) + \Delta \sum_{\vec{i},\sigma}(-1)^{s_{\vec i}} \hat n_{\vec i\sigma},\nonumber 
\end{eqnarray}
where $\hat c^{\dagger}_{\vec{i}\sigma}$ ($\hat c^{}_{\vec{i}\sigma}$) are the fermionic creation (annihilation) operators at site $\vec {i}$, with (pseudo-)spin $\sigma=\uparrow,\downarrow$, and $\hat n_{\vec i\sigma}= \hat c^{\dagger}_{\vec{i}\sigma} \hat c^{}_{\vec{i}\sigma}$ is the corresponding (pseudo-)spin site-occupation operator. The nearest-neighbor, $\langle {\vec {i\,j}} \rangle$, and next-nearest neighbor (NNN), $\langle\langle {\vec {i\,j}} \rangle\rangle$, hopping parameters are denoted by $t$ and $t^\prime$, respectively; the strength of the on-site \textit{attractive} interaction by $U<0$, and the strength of the staggered potential by $\Delta$ [$s_{\vec i}$ is either 0 or 1 depending on the sublattice to which ${\vec i}$ belongs to, see Fig.~\ref{fig:cartoon_bands_dos}(a)]. In what follows, we focus on an average filling of one electron per site $N \equiv N_\uparrow+N_\downarrow = V$, with $N_\uparrow \equiv\sum_{\vec{i}}\langle \hat n_{\vec i\uparrow}\rangle = N_\downarrow \equiv \sum_{\vec{i}}\langle \hat n_{\vec i\downarrow}\rangle$, and $V=2L^2$ being the number of lattice sites [$L$ is the number of unit cells in each direction, see Fig.~\ref{fig:cartoon_bands_dos}(a)]. We set $t=1$ as the energy scale, but write $t$ whenever it is helpful in equations. 

Compared to the model studied experimentally in Ref.~\cite{Messer2015}, Eq.~\eqref{eq:Hamiltonian} contains only an extra set of terms, namely, the NNN hoppings, which describe hoppings between sites that belong to the same sublattice. As will become apparent below, those terms introduce nontrivial changes to the solely nearest-neighbors model, and they can be introduced in experiments via a modulated shaking of the optical lattice~\cite{Jotzu2014}.

\begin{figure}[!t] %%% FIG2
\includegraphics[width=0.98\columnwidth]{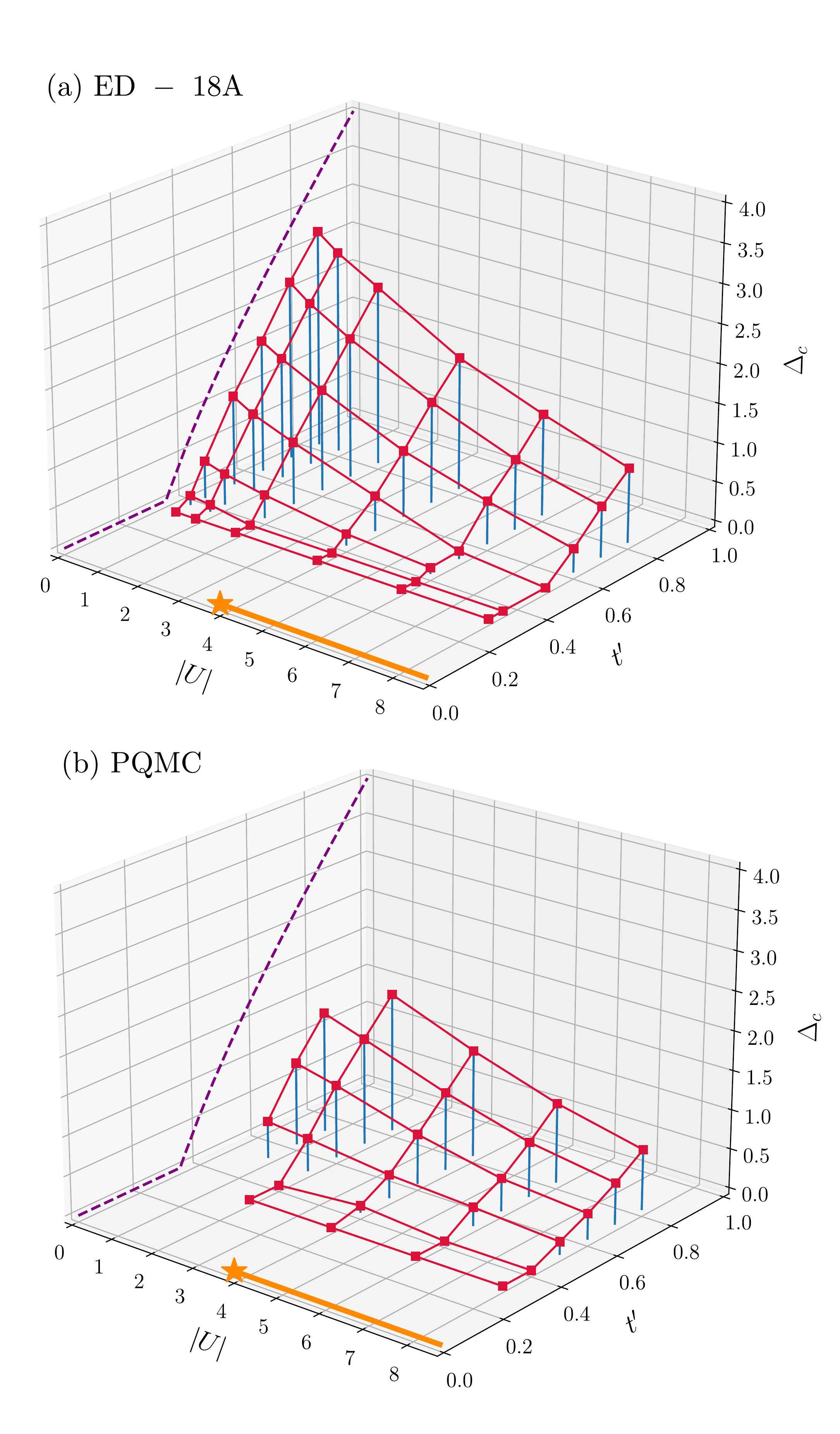}
    \vspace{-0.3cm}
    \caption{(a) Phase diagram of Hamiltonian~\eqref{eq:Hamiltonian} obtained via exact diagonalization in an 18-site lattice; similar results for a smaller lattice are reported in Appendix~\ref{sec:16B}. The dashed line in the $U/t=0$ plane shows the boundary between the metallic and band-insulating phases obtained analytically. For nonzero $U$, the surface formed by connecting the points (which report $\Delta_c$) delimits the insulating ($\Delta>\Delta_c$) and superconducting ($\Delta<\Delta_c$) phases. (b) The phase diagram obtained from PQMC calculations in larger lattices, using a scaling ansatz for results obtained in lattices with $L=6, 9, 12$, and 15. In both panels, the onset of the supersolid regime (or an SU(2) antiferromagnetic Mott insulator in the repulsive language~\cite{Lee2009}) at $t^\prime = 0$ is marked by a star.}
  \label{fig:phase_diagram}
\end{figure}

To understand the interplay of the different terms in Eq.~\eqref{eq:Hamiltonian}, it is useful to analyze several of its limiting regimes. First, let us discuss the noninteracting ($U=0$) limit, in which the Hamiltonian is diagonalizable in $\vec{k}$ space resulting in the following two bands:
\begin{equation}
    E_{\vec k} = -t^\prime f({\vec k}) \pm \sqrt{\Delta^2 + t^2\left[3+f({\vec k})\right]},
\label{eq:non_int_dispersion}
\end{equation}
with $f({\vec k}) = 2 \cos(\sqrt{3}k_y) + 4 \cos\left(\frac{3}{2}k_x\right) \cos\left(\frac{\sqrt{3}}{2} k_y \right)$. The first question one can ask is how large $\Delta$ needs to be for the ground state at half filling to transition between the semi-metal ground state for $\Delta=0$ to the band insulator for $\Delta\gg t$. The answer to that question depends on the NNN hopping amplitude $t^\prime$. For $t^\prime=0$, the Dirac cones of the upper and lower bands, touching at the high symmetry points K and K$^\prime$ of the first Brillouin zone, split, generating a direct gap ($=2\Delta$) for any nonzero value of $\Delta$. After including a nonzero $t^\prime$, the direct gap is still obtained so long as $t^\prime\leq t^\prime_c= t/3$ [see Fig.~\ref{fig:cartoon_bands_dos}(b)]. As $t^\prime$ increases crossing $t^\prime_c$, the minimum of the upper band moves from the K,K$^\prime$ points to the $\Gamma$ point. For $t^\prime>t^\prime_c$, the gap becomes indirect (K,K$^\prime\leftrightarrow\Gamma$) and the amplitude of the staggered potential $\Delta_c$ needed to turn the system into a band-insulator becomes non-vanishing [see Fig.~\ref{fig:cartoon_bands_dos}(c)]. Specifically, $\Delta_c=\frac{9t^{\prime 2}-t^2}{2t^\prime}$, as depicted by the dashed-dotted line in the $U=0$ plane in Fig.~\ref{fig:phase_diagram}.

A second important limit that can be promptly understood is the $t^\prime=0$ case for $U\neq0$. In this regime, see Ref.~\cite{Mondaini2015} for a similar discussion in the context of the square lattice, a particle-hole transformation on only one of the components of the (pseudo-)spin, say, the $\downarrow$-component, $\hat c_{{\vec i},\downarrow}\leftrightarrow(-1)^{s_{\vec i}}\hat c_{{\vec i},\downarrow}^\dagger$ ($s_{\vec i}$ is either 0 or 1, depending on the sublattice to which ${\vec i}$ belongs), maps the original attractive Hubbard model onto the \textit{repulsive} Hubbard model in the presence of a staggered Zeeman field, i.e., $\Delta\sum_i (-1)^{s_{\vec i}}(\hat n_{\vec{i}\uparrow} + \hat n_{\vec{i}\downarrow}) \leftrightarrow \Delta \sum_i(-1)^{s_{\vec i}} (\hat n_{\vec{i}\uparrow} - \hat n_{\vec{i}\downarrow})$. Since a nonzero staggered field ($\Delta\neq0$) in the repulsive Hubbard model in the honeycomb lattice explicitly breaks SU(2) symmetry, the antiferromagnetic Mott insulator for $U_c \gtrsim 3.8$~\cite{Paiva05, Meng10, Sorella2012, Assaad2013, Toldin2015, Otsuka2016} becomes an $S_z$ antiferromagnet. Recalling the mapping between magnetic and charge/pair degrees of freedom in the repulsive and attractive Hubbard models under the particle-hole transformation~\cite{Lee2009}, 
\begin{eqnarray}
    2 \hat S^z_{\vec{i}} = \hat n_{\vec{i}\uparrow} - \hat n_{\vec{i}\downarrow} &\leftrightarrow& \hat n_{\vec{i}\uparrow} + \hat n_{\vec{i}\downarrow} = \hat n_{\vec{i}} \nonumber \\
    \hat S^+_{\vec{i}} = \hat c_{\vec{i}\uparrow}^\dag \hat c_{\vec{i}\downarrow}^{} &\leftrightarrow& \hat c_{\vec{i}\uparrow}^\dag \hat c_{\vec{i}\downarrow}^\dag = \hat \Delta_{\vec i}^\dag \nonumber \\
    \hat S^-_{\vec{i}} = \hat c_{\vec{i}\downarrow}^\dag \hat c_{\vec{i}\uparrow}^{} &\leftrightarrow& \hat c_{\vec{i}\downarrow} \hat c_{\vec{i}\uparrow} = \hat \Delta_{\vec{i}},
\end{eqnarray}
one finds that any nonvanishing value of the staggered potential $\Delta$ in the attractive Hubbard model breaks the supersolid state at $|U|>|U_c|$, leading to an insulating ground state with a different $\langle \hat n_{\vec{i}}\rangle$ in the two sublattices that make the bipartite honeycomb lattice. An important point to keep in mind is that this difference in the site occupations in the sublattices results from having the staggered potential. It is not the result of a spontaneous symmetry-breaking process, i.e., the insulating ground state is a Mott insulator that does not break symmetries of the model. For $|U| < |U_c|$, as for $U=0$ and $|U|>|U_c|$, we expect that any non-vanishing value of the staggered potential $\Delta$ leads to an insulating ground state.

For $t^\prime>t^\prime_c$ and $\Delta=0$, we mentioned before that the ground state at $U=0$ is a metal. This means that adding attractive on-site interactions $U<0$ results in a superconducting state. Hence, in the regime with $t^\prime>t^\prime_c$ and $U<0$, a nonzero value $\Delta_c$ of the staggered potential is needed to drive the SIT. In the absence of unbiased analytical techniques to tackle this regime for arbitrary values of $U$, $\Delta$, and $t^\prime$, we carry out numerical calculations to obtain the phase diagram of Hamiltonian~\eqref{eq:Hamiltonian} as characterized by the surface $\Delta_c(U,t^\prime)$. A compilation of the results obtained using Krylov-based exact diagonalization~\cite{Petsc,Slepc} in a lattice with $V = 2L^2 = 18$ sites is shown in Fig.~\ref{fig:phase_diagram}(a). This lattice, which has $L=3$ and is depicted in Fig.~\ref{fig:cartoon_bands_dos}(a), contains the Brillouin zone corner as a valid momentum point, i.e., it captures the effects of low-energy excitations about the high-symmetry K points. This makes it optimal to determine the phase diagram within the lattice sizes that we can study using exact diagonalization~\footnote{See, e.g., Ref.~\cite{ge_rigol_17} for a discussion of finite-size effects in topological transitions in commensurate vs. incommensurate lattices.}. Qualitatively similar results were obtained on a 16-sites lattice; see Appendix~\ref{sec:16B}.

Much larger lattices can be studied by means of projective quantum Monte Carlo (PQMC) simulations. Within this approach, the ground state $|\Psi_0\rangle$ is obtained projecting a trial wave function $|\Psi_T\rangle$, via $|\Psi_0\rangle = \lim_{\Theta\to\infty} e^{-\Theta \hat H}|\Psi_T\rangle$, where $\Theta$ is a projector parameter. This approach works provided that the overlap between the trial wave function and the ground-state of Hamiltonian~\eqref{eq:Hamiltonian} is nonzero, $\langle \Psi_0|\Psi_T\rangle\neq 0$, and that $|\Psi_0\rangle$ is nondegenerate~\cite{Assaad2002}. The expectation values of operators $\hat O$ in the ground state can then be written as
\begin{equation}
    \langle \hat O\rangle = \frac{\langle \Psi_0 |\hat O |\Psi_0\rangle}{\langle \Psi_0|\Psi_0 \rangle} = \lim_{\Theta \to \infty}\frac{\langle \Psi_T|e^{-\Theta\hat H} \hat O e^{-\Theta\hat H}|\Psi_T\rangle}{\langle \Psi_T|e^{-2\Theta\hat H}|\Psi_T \rangle}.
\end{equation}
For our simulations, we considered two trial wave functions, the half-filled Fermi sea of the noninteracting part of the Hamiltonian and a Hartree-Fock solution. We found the latter to converge more quickly with increasing $\Theta$ so all results reported are obtained with that trial wave function for $\Theta t = 40$, which is sufficiently large for the convergence of the expectation values of all observables studied here. Since our PQMC calculations are carried out within the canonical ensemble for $N_\uparrow = N_\downarrow$, they do not suffer from the sign problem~\cite{Loh1990, Assaad2002, Iglovikov2015, Mondaini2022}. The phase diagram $\Delta_c(U,t^\prime)$ extracted from the PQMC simulations of lattices with up to $L = 15$ is reported in Fig.~\ref{fig:phase_diagram}(b). It is similar to the one obtained using exact diagonalization in small lattices.

In the next section, we discuss in detail how the phase diagrams reported in Fig.~\ref{fig:phase_diagram} are obtained using exact diagonalization and PQMC simulations.

\section{Phase diagram calculations}\label{sec:phasediagram}

Before explaining in detail how we locate the phase transition using exact diagonalization and PQMC simulations and how we probe its universality class, let us briefly comment on the overall structure of the phase diagrams in Fig.~\ref{fig:phase_diagram}. Increasing attractive interactions results in smaller Cooper pairs: In the limit $|U| \to \infty$ they fit in a site, becoming hardcore bosons~\cite{Micnas1990}. Consequently, the magnitude of the staggered potential needed to transition between the superfluid and insulating ground states decreases as the increasingly local fermionic pairs at stronger interactions become pinned at the $-\Delta$ sites in the lattice. On the other hand, the NNN hopping terms act to counteract the pinning promoted by the staggered potential, so increasing the magnitude of $t^\prime$ results in the need for a stronger staggered potential to induce the superfluid-insulator transition. Notably, the phase diagram also indicates that the transition at nonvanishing interactions is smoothly connected to the noninteracting metal-band insulator one, marked by the dash-dotted lines at $U = 0$ in both panels of Fig.~\ref{fig:phase_diagram}. We also note that, for values of $t^\prime > t/3$, the noninteracting regime at $\Delta=0$ no longer corresponds to a semi-metal but rather to a metal with a finite density of states at the Fermi level. Hence, the inclusion of attractive interactions results in pairing and superconductivity.

\subsection{SIT in exact diagonalization calculations}\label{sec:ed}

In our exact diagonalization calculations, to identify the critical value $\Delta_c$, for any given value of $U$ and $t^\prime$, we use the fidelity susceptibility~\cite{Zanardi06, CamposVenuti07, Zanardi07, You2007},
\begin{equation}
    g_\Delta = \frac{2}{V} \frac{1 - |\langle \Psi_0(U,t^\prime,\Delta) | \Psi_0(U,t^\prime,\Delta + \delta\Delta) \rangle|}{(\delta\Delta)^2},
\end{equation}
for which one needs to compute the fidelity of ground-state wave functions for slightly different Hamiltonian parameters; we modify the staggered potential by $\delta\Delta = 10^{-3}$ in our calculations. Continuous phase transitions can be identified by large peaks that appear as a critical point is crossed~\cite{Yang07, Jia11, Mondaini2015, Yi2021}. 

\begin{figure}[!t] %%% FIG3
    \includegraphics[width=0.98\columnwidth]{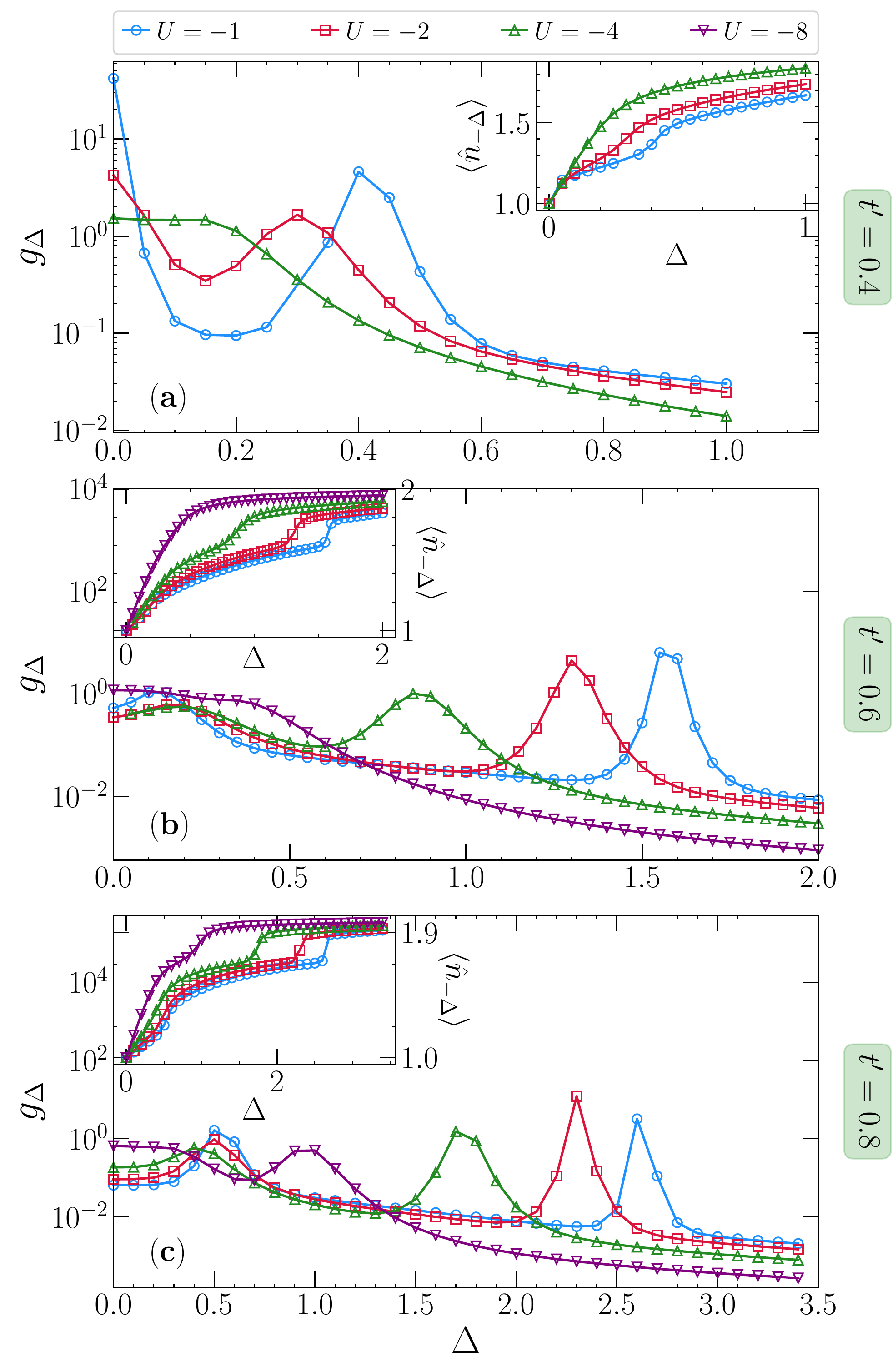}
    \vspace{-0.3cm}
    \caption{Fidelity susceptibility computed using exact diagonalization in the 18-site cluster shown in Fig.~\ref{fig:cartoon_bands_dos}(a) for different values of $U$, when (a) $t^\prime = 0.4$, (b) $t^\prime = 0.6$, and (c) $t^\prime = 0.8$. Insets: Corresponding occupation of the $-\Delta$ sites.}
  \label{fig:ED_g_nud}
\end{figure}

In Fig.~\ref{fig:ED_g_nud}, we show results for $g_\Delta$ vs $\Delta$ for different values of $U$, when $t^\prime=0.4$ [Fig.~\ref{fig:ED_g_nud}(a)], $t^\prime=0.6$ [Fig.~\ref{fig:ED_g_nud}(b)], and $t^\prime=0.8$ [Fig.~\ref{fig:ED_g_nud}(c)]. For $t^\prime=0.4$ in Fig.~\ref{fig:ED_g_nud}(a), one can see that the curves for different values of $U$ exhibit sharp peaks (notice the logarithmic scale in the $y$ axis) in $g_\Delta$ at $\Delta=0$ and nonzero values of $\Delta$. The $\Delta=0$ peaks signal changes in the superconducting state resulting from introducing the staggered potential. [With increasing $t^\prime$, see Figs.~\ref{fig:ED_g_nud}(b) and~\ref{fig:ED_g_nud}(c), those peaks can be seen to move to nonzero values of $\Delta$.] The peaks at nonzero values of $\Delta$ in Fig.~\ref{fig:ED_g_nud}(a) signal the SIT, $\Delta_c$. As advanced, for $t^\prime$ fixed, increasing $|U|$ results in a decrease of $\Delta_c$. Figures~\ref{fig:ED_g_nud}(b) and~\ref{fig:ED_g_nud}(c) show that increasing $t^\prime$, at any given value of $U$, results in an increase of $\Delta_c$. The compilation of such peak locations in the space of parameters $(U,t^\prime)$ leads to the phase diagram in Fig.~\ref{fig:phase_diagram}(a).

The insets in Fig.~\ref{fig:ED_g_nud} show that at $\Delta_c$ the occupation of the sites with chemical potential $-\Delta$, in short, the $-\Delta$ sites, exhibits a rapid increase after which it nearly saturates to the maximal possible value $\langle \hat n_{\vec{i}}\rangle=2$. The derivatives of the site occupations, as well as of the double occupancy
\begin{equation}
    D_{\vec{i}} = \langle \hat n_{\vec{i}\uparrow} \hat n_{\vec{i}\downarrow}\rangle,
\end{equation}
and the kinetic energy per site
\begin{equation}
    {\cal K} = -\frac{1}{V} \sum_{{\vec {ij}},\sigma} t_{\vec{i}\vec{j}}\langle\hat c^{\dagger}_{\vec{i}\sigma} \hat c^{}_{\vec{j}\sigma}\rangle,
\end{equation}
with respect to $\Delta$ was used to track the SIT in a related model~\cite{Mondaini2015}, and can be measured in ultracold gases experiments~\cite{Messer2015}. It was recently shown that studying the dynamics of local quantities after quantum quenches also allows one to locate quantum phase transitions~\cite{haldar_mallayya_21}.

\begin{figure}[!t] %%% FIG4
    \includegraphics[width=0.98\columnwidth]{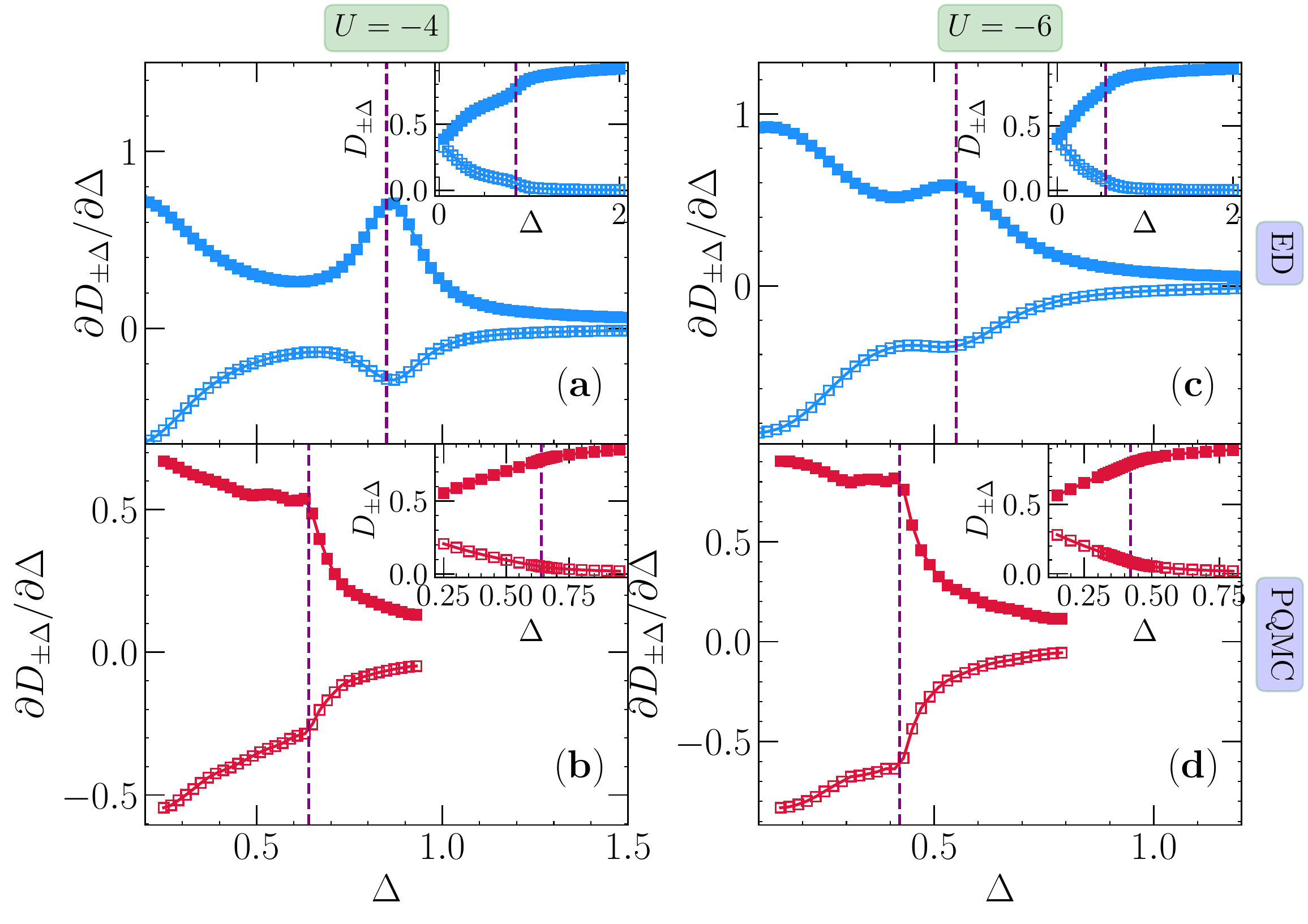}
    \vspace{-0.3cm}
    \caption{Numerical derivative of the double occupancy with respect to $\Delta$ for (a), (b) $U = -4$ and (c), (d) $U = -6$; the insets display the double occupancy before differentiation. Empty (filled) symbols denote the double occupancy in the $+\Delta$-sites [$-\Delta$-sites]. (a), (c) Results obtained using exact diagonalization in a lattice with $L=3$. (b), (d) Results obtained using PQMC in a lattice with $L=15$. The vertical dashed lines depict $\Delta_c$ as identified using exact diagonalization [(a), (c)] and PQMC [(b), (d)]. All these results were obtained for $t^\prime = 0.6$.}
  \label{fig:PQMC_local_qtts}
\end{figure}

In Figs.~\ref{fig:PQMC_local_qtts}(a) and~\ref{fig:PQMC_local_qtts}(c), we show exact diagonalization results for $\partial D/\partial \Delta$ vs $\Delta$ in the two sublattices for $t'/t=0.6$, when $U/t=-4$ [Fig.~\ref{fig:PQMC_local_qtts}(a)] and $U/t=-6$ [Fig.~\ref{fig:PQMC_local_qtts}(c)]. The corresponding evolutions of the double occupancies are shown in the insets. The vertical dashed lines depict $\Delta_c$ as identified using the fidelity susceptibility. They coincide with the values of $\Delta$ at which $\partial D/\partial \Delta$ exhibit local extrema. In Figs.~\ref{fig:PQMC_local_qtts}(b) and~\ref{fig:PQMC_local_qtts}(d), we show results for $\partial D/\partial \Delta$ vs $\Delta$ obtained using PQMC in much larger lattices for the same parameters as in Figs.~\ref{fig:PQMC_local_qtts}(a) and~\ref{fig:PQMC_local_qtts}(c), respectively. The vertical dashed lines depict $\Delta_c$, identified using PQMC (as explained in Sec.~\ref{sec:sitpqmc}). In those larger system sizes, $\partial D/\partial \Delta$ exhibits a kink at $\Delta_c$, after which it decreases rapidly. A more prominent kink, also signaling $\Delta_c$, is seen in the derivative of the kinetic energy, see Appendix~\ref{sec:kin}. Comparing the results obtained using PQMC and exact diagonalization, one notices that the values of $\Delta_c$ are different, and that the differences decrease with increasing $|U|$. This is expected as the exact diagonalization results are affected by finite-size effects, and finite-size effects are stronger in the weakly interacting regime in which the pair sizes are larger.

In the context of ultracold fermionic atoms experiments, we note that double occupancy was originally used to identify the Mott insulating regime~\cite{Jordens2008}. It is currently used, together with other local observables, to experimentally characterize fermionic systems with single-site resolution~\cite{Greif2016, Boll2016, Cheuk2016}.

\subsection{SIT in PQMC calculations}\label{sec:sitpqmc}

The critical points in our PQMC simulations are identified using the scaling of the pair structure factor 
\begin{equation}
    P_s = \frac{1}{V}\sum_{\vec{i},\vec{j}} \langle \hat \Delta_{\vec{i}}^{}\hat \Delta_{\vec{j}}^\dagger\rangle,
\end{equation}
where $\hat \Delta_{\vec{i}}^{} \equiv \hat c_{\vec{i}\uparrow}^{} \hat c_{\vec{i}\downarrow}^{}$ ($\hat \Delta_{\vec{i}}^\dagger = \hat c_{\vec{i}\downarrow}^\dagger\hat c_{\vec{i}\uparrow}^\dagger$) is the pair annihilation (creation) operator at site $\vec{i}$. Long-range order in the superconducting phase means that $P_s$ is extensive in $V$, but the way such behavior is approached when decreasing the staggered potential from the insulating phase is controlled by the critical exponents.

In Ref.~\cite{Mondaini2015}, a scaling ansatz for $P_s$ was discussed for a similar model in the square lattice. Next, we summarize the arguments presented there. To start, we note that, in the strongly interacting (large-$|U|$) regime, second-order perturbation theory shows that our model effectively becomes a model of repulsive hardcore bosons in the presence of a staggered potential~\cite{Robaszkiewicz1981, Micnas1990, Mondaini2018}. The creation and annihilation operators of hardcore bosons are related to the annihilation and creation of pairs, respectively, $\hat b_{\vec{i}} = \hat \Delta_{\vec{i}}^\dagger$ and $\hat b_{\vec{i}}^\dagger = \hat \Delta_{\vec{i}}$. In the absence of NNN hoppings and interactions, such a hard-core boson model was studied in Refs.~\cite{Hen2009, Hen2010} in square and cubic lattices. The transition between the superfluid and the Mott-insulating state, driven by the staggered potential, was shown to belong to the $(d+1)$-XY universality class, which is the same universality class of the superfluid--Mott-insulator transition in the Bose-Hubbard model at fixed integer site occupancy~\cite{fisher_89}. 

The previously mentioned operator mapping brings a direct analogy: The $s$-wave pair structure factor translates into the zero-momentum occupancy for hardcore bosons in the effective model, $n_{\vec{k}=0}=(1/V)\sum_{\vec{i},\vec{j}}\langle \hat b_{\vec{i}}^\dagger \hat b_{\vec{j}} \rangle$, which is known to diverge when approaching the transition from the normal side as $n_{\vec{k}=0} \sim \xi^{1-\eta}$~\cite{Pollet2010, Carrasquilla2012}, where $\xi$ is the correlation length and $\eta=0.0381(2)$~\cite{Campostrini2006} the anomalous scaling dimension. In a finite system, this relation implies that the `condensate fraction' $f_0 \equiv n_{\vec{k}=0}/N_{\rm pairs}$ (with $N_{\rm pairs} = V/2 = L^2$) vanishes at the critical point as $f_0 \sim L^{-(1+\eta)}$~\cite{Pollet2010, Carrasquilla2012}. A scaling ansatz thus naturally follows as $f_0 L^{1+\eta} = g(|\Delta-\Delta_c|L^{1/\nu})$, with $\nu=0.6717(1)$ (see Ref.~\cite{Campostrini2006}) the critical exponent related to the divergence of the correlation length at $\Delta\to\Delta_c$. Translating it to the original fermionic model, we get
\begin{equation}
    \left(\frac{P_s}{N_{\rm pairs}}\right)L^{1+\eta} = g(|\Delta-\Delta_c|L^{1/\nu})\ .
    \label{eq:ansatz}
\end{equation}

\begin{figure}[!tb] %%% FIG5
    \includegraphics[width=0.95\columnwidth]{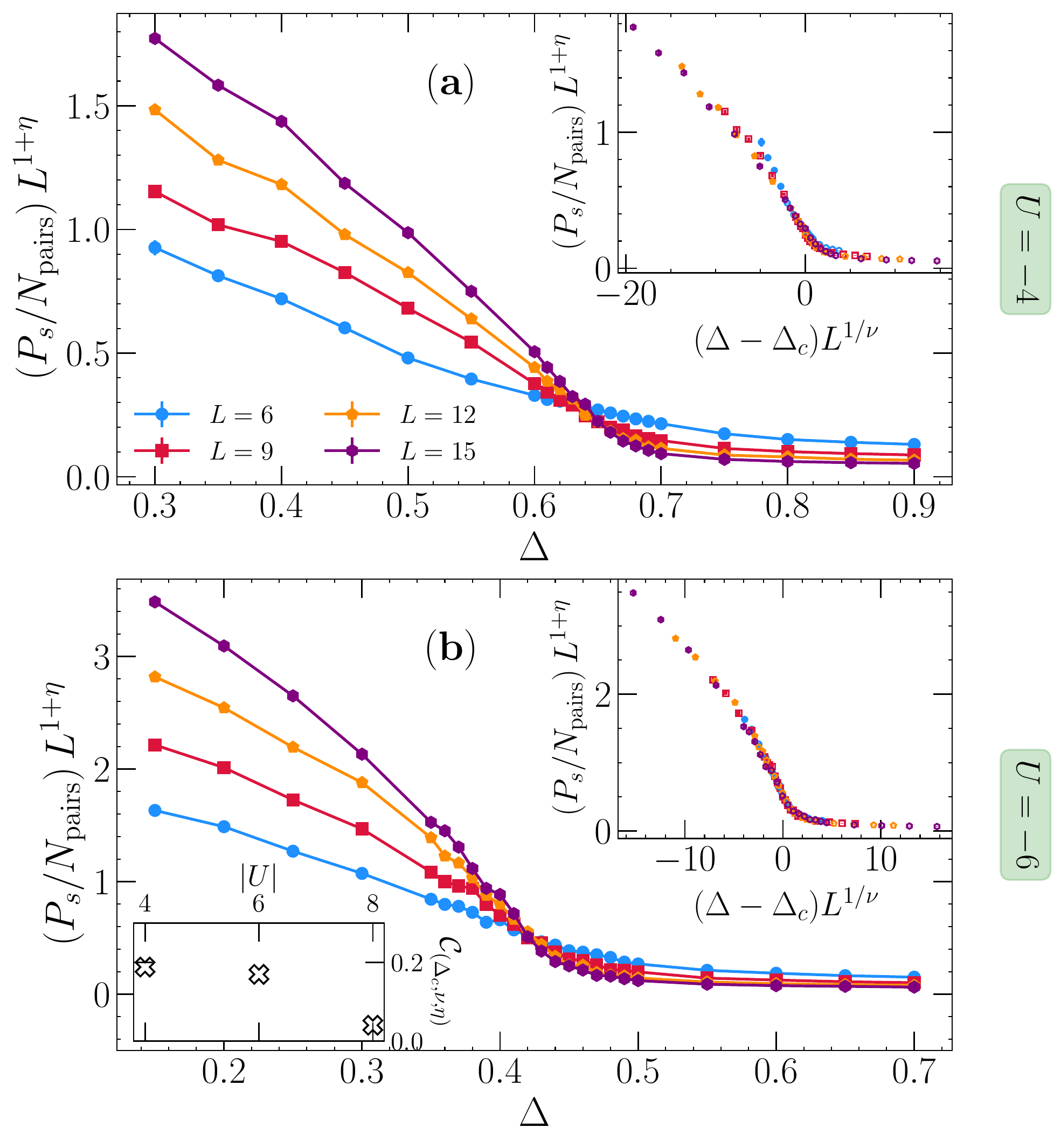}
    \vspace{-0.3cm}
    \caption{Scaled pair structure factor versus $\Delta$, for fixed $t^\prime=0.6$ and two values of $U$, (a) $U = -4$ and (b) $-6$. The upper right insets display the corresponding scaling collapse following the ansatz~\eqref{eq:ansatz}, with $\Delta_c = 0.637(1)$ and $0.420(1)$ for $U=-4$ and $-6$, respectively. The lower inset in (b) depicts the $|U|$ dependence of the cost function used to quantify the quality of data collapse; see text.}
  \label{fig:PQMC_scaling}
\end{figure}
    
In Fig.~\ref{fig:PQMC_scaling}, we show the scaled pair structure factor for two on-site attractive interaction strength values at $t^\prime = 0.6$. Crossings of the curves for different system sizes can be seen to occur at specific values of $\Delta$ that depend on $U$; these crossings allow us to identify $\Delta_c$ using our PQMC calculations. The results in Fig.~\ref{fig:PQMC_scaling} show that the strength of the staggered potential needed to induce the SIT decreases with increasing the strength of the on-site attractive interactions. The upper right insets in Figs.~\ref{fig:PQMC_scaling}(a) and \ref{fig:PQMC_scaling}(b) display the scaling collapse according to the ansatzin Eq.~\eqref{eq:ansatz}.

The analysis in Fig.~\ref{fig:PQMC_scaling} confirms that the SIT in our model belongs to the 3$d$-XY universality class even if one is not in a regime that can be described using an effective hard-core boson Hamiltonian. Using the cost function ${\cal C}_{(\Delta_c, \nu, \eta)} = (\sum_j |y_{j+1} - y_j|)/(\max\{y_j\} - \min\{y_j\})-1$~\cite{Suntajs2020, Mondaini2022b}, where $y_j$ are the values of $(P_s/N_{\rm pairs})L^{1+\eta}$ ordered according to their corresponding $(\Delta-\Delta_c)L^{1/\nu}$'s, allows us to quantify the quality of the scaling collapse. We show results for this cost function in the lower inset in Fig.~\ref{fig:PQMC_scaling}(b). They reveal that, for the system sizes considered, the collapse improves with increasing $|U|$. This is expected to be the result of a reduction of the finite-size effects as the size of the Copper pairs decreases. A compilation of the values of $\Delta_c$ obtained via the crossing of the scaled pair structure factor in the space of parameters $(U,t^\prime)$ gives rise to the phase diagram in Fig.~\ref{fig:phase_diagram}(b).

\section{Charge excitations}\label{sec:excitations}

The 3$d$-XY universality class of the transition advances that the lowest energy charge excitations across the transition are bosonic~\cite{Mondaini2015}. With increasing $\Delta$ in the insulating phase, the lowest-energy charge excitations must cross over from bosonic to fermionic (at $\Delta=\infty$ they are fermionic for any finite $U$). 

In this section, we study the nature of the lowest-energy charge excitations across the SIT and their behavior with increasing $\Delta$ in the insulating phase. For this, we compute the one-particle ($m=1$, fermionic) and two-particle ($m=2$, bosonic) gaps~\cite{Dodaro2017, Junemann2017},
\begin{equation}
    \delta^{(m)} = E_0(N + m) + E_0(N - m) - 2 E_0(N),
    \label{eq:m_gap}
\end{equation}
where $E_0(x)$ is the ground-state energy with $x$-fermions. For $m=2$, we always add/remove one fermion with (pseudo-)spin $\uparrow$ and one with (pseudo-)spin $\downarrow$, namely, a pair. Hence, in what follows, we call the $m=2$ gap the pair gap.

The one-particle and pair gaps in Eq.~\eqref{eq:m_gap} can be straightforwardly computed using exact diagonalization. Similarly, since for $m=2$ we have that $N_\uparrow=N_\downarrow$ so there is no sign problem, the pair gap in Eq.~\eqref{eq:m_gap} can also be computed using PQMC simulations in much larger system sizes. PQMC runs into the sign problem for $m=1$ in Eq.~\eqref{eq:m_gap}, because $N_\uparrow\neq N_\downarrow$. Hence, we follow a different approach to compute the one-particle gap. We probe the decay of the appropriate time-displaced correlation function~\cite{Meng10}. Specifically, the one-particle gap is extracted using the imaginary-time $\tau$ displaced Green's functions $G_+({\bf k}, \tau) = \langle \hat c_{{\bf k}, \sigma}(\tau)\hat c^{\dagger}_{{\bf k},\sigma}(0)\rangle$ and $G_-({\bf k}, \tau) = \langle \hat c^\dagger_{{\bf k}, \sigma}(\tau)\hat c_{{\bf k},\sigma}(0)\rangle$, with $\hat c^\dagger_{{\bf k}, \sigma}(\tau) \equiv e^{\tau \hat H}c^{\dagger}_{{\bf k},\sigma}(0) e^{-\tau \hat H}$, which describe the particle and hole excitations with respect to the Fermi energy, respectively. At large $\tau$'s, they decay as $G_\pm({\bf k}, \tau) \propto e^{-\tau\delta_\pm^{(1)}({\bf k})}$. By comparing the two branches over different momenta ${\bf k}$, the one-particle gap is obtained as $\delta^{(1)} = \min_{\bf k}[\delta_+^{(1)}({\bf k})] + \min_{\bf k}[\delta_-^{(1)}({\bf k})]$ (see Appendix~\ref{sec:G_tau} for further details about this analysis).

\begin{figure}[!tb] %%% FIG6
  \includegraphics[width=0.85\columnwidth]{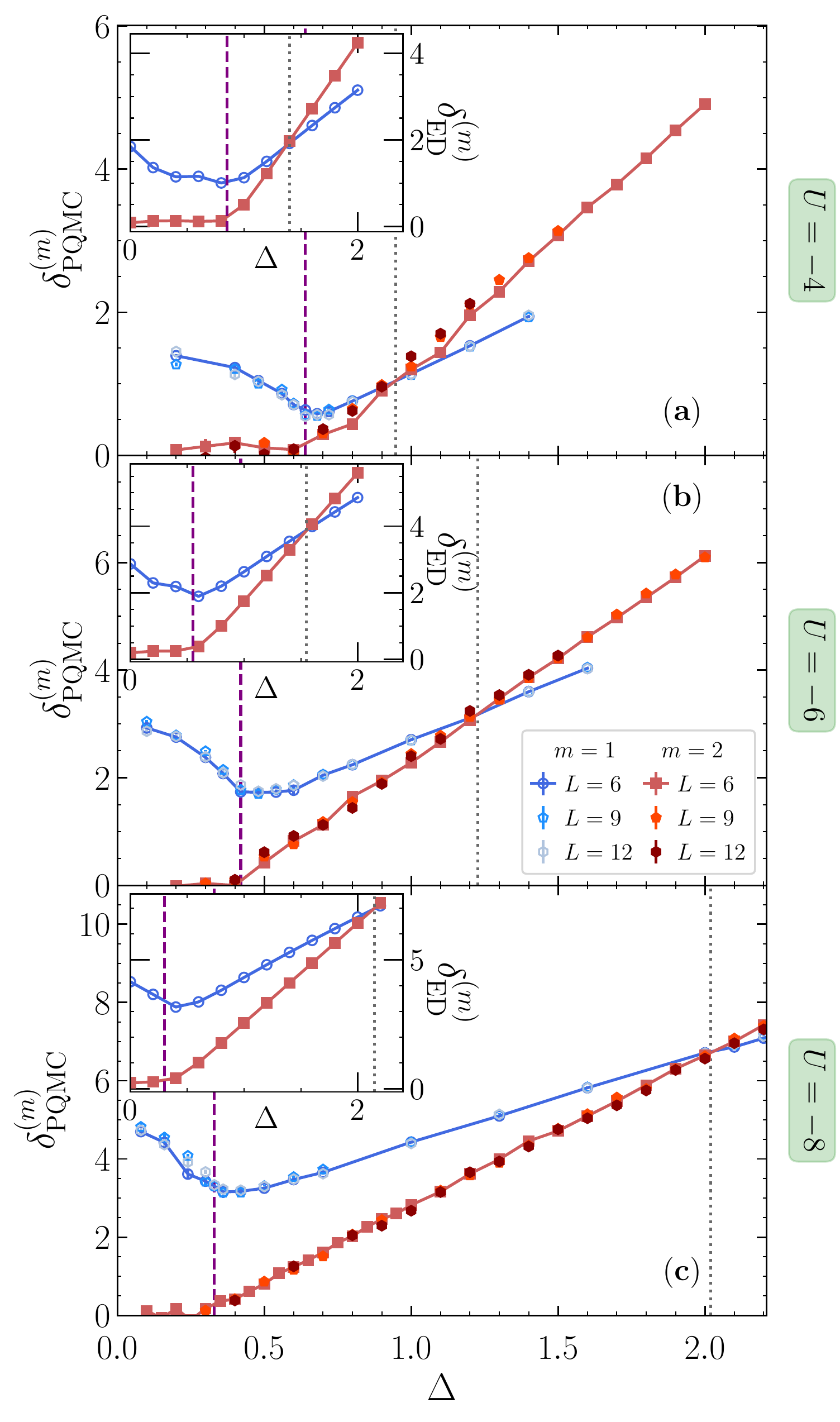}
  \vspace{-0.3cm}
  \caption{(a)--(c) The one-particle and pair gap dependence on the staggered potential strength $\Delta$ in both PQMC calculations for $L=6$, 9, and 12 (main panels) and ED calculations for $L=3$ (insets). We show results for $t^\prime = 0.6$, and (a) $U=-4$, (b) $U=-6$, and (c) $U=-8$. In all panels, the vertical dashed lines depict the SIT location as obtained using the corresponding computational technique, while the vertical dotted lines show the value of $\Delta$ at the crossing of the gaps.}
  \label{fig:gaps}
\end{figure}

Focusing on $t^\prime=0.6$, Figs.~\ref{fig:gaps}(a)--\ref{fig:gaps}(c) display the one-particle and pair gaps ($m=1$ and $2$) obtained using PQMC in lattices with $L=6$, 9, and 12 (main panels) and using exact diagonalization in a lattice with $L=3$ (insets). The PQMC and exact diagonalization results are qualitatively similar. The pair gap vanishes in the superconducting phase. It then becomes nonzero, with a magnitude that increases with increasing $\Delta$, once the SIT (marked by the dashed line) is crossed. On the other hand, the one-particle gap is nonvanishing in both the superconducting and insulating phases and exhibits a minimum at the SIT. Hence, as advanced, the pair gap is smaller than the one-particle gap in both the superconducting and insulating phases across the SIT. With increasing $\Delta$, those gaps cross in the insulating phase at a value of $\Delta$ that, deep in the strongly interacting regime, increases with increasing $|U|$. The PQMC results for different system sizes show that finite-size effects are small in the gaps computed for those system sizes, and even more so as $|U|$ and $\Delta$ increase. This suggests that our PQMC estimation of the crossing points does not suffer from significant finite-size effects.

\begin{figure}[!tb] %%% FIG7
    \includegraphics[width=0.9\columnwidth]{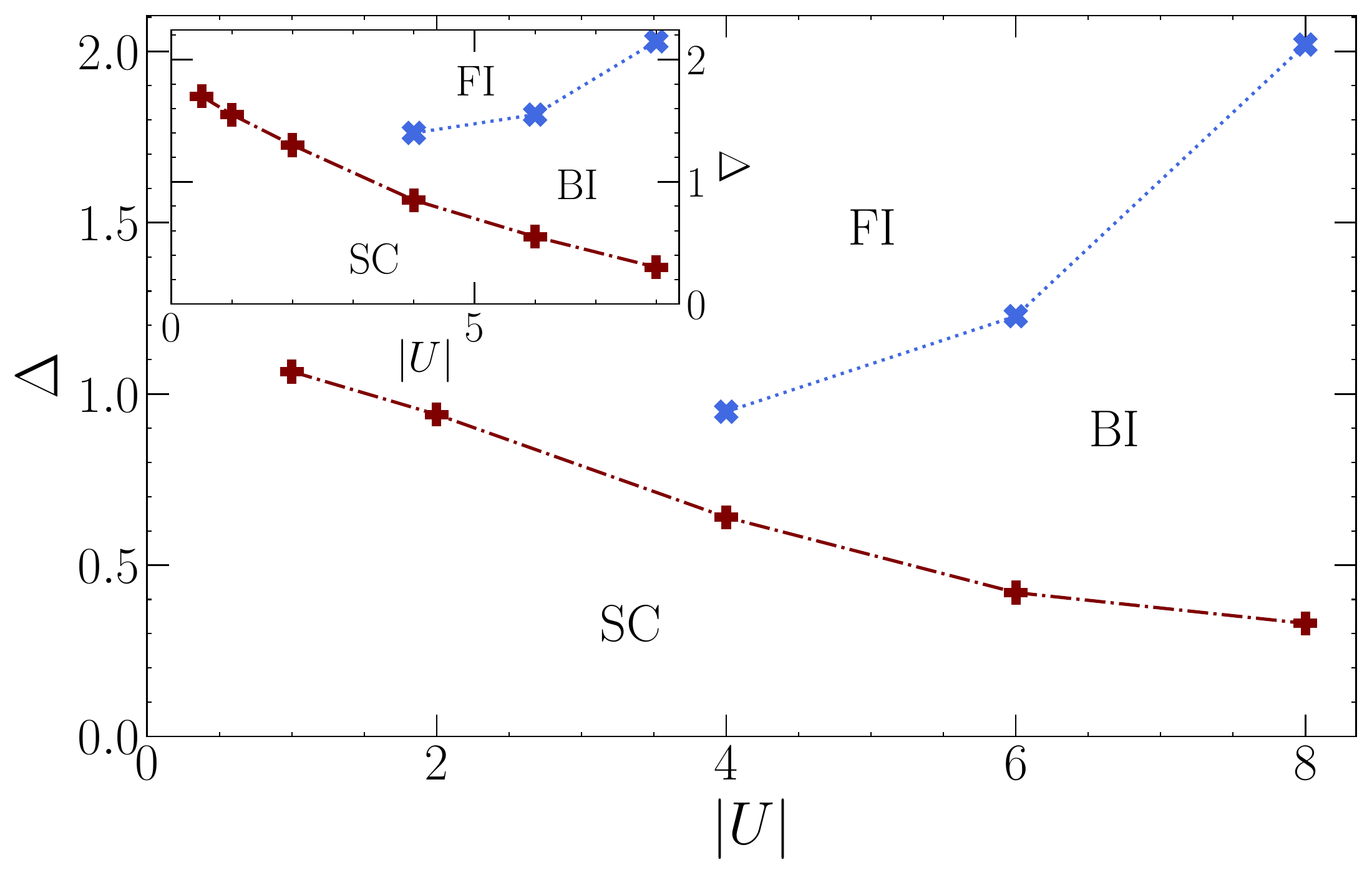}
    \vspace{-0.3cm}
    \caption{Phase diagram in the ($U,\Delta$) plane for $t^\prime=0.6$ obtained via PQMC (main panel) and exact diagonalization in a lattice with $L=3$ (inset). The lower curve depicts the location of the SIT; between the superconducting phase (SC) and the insulating phase with lowest-energy charge excitations that are bosonic (BI). The upper curve depicts the crossover location in the insulating phase between the regime in which the lowest-energy charge excitation is bosonic (BI) and fermionic (FI).}
  \label{fig:exc_diag}
\end{figure}

Compiling results like those depicted in Fig.~\ref{fig:gaps} allows us to obtain exact diagonalization (inset in Fig.~\ref{fig:exc_diag}) and PQMC (main panel in Fig.~\ref{fig:exc_diag}) phase diagrams in the ($U,\Delta$) plane for $t^\prime=0.6$. They highlight that, with increasing $|U|$, there is an increase in the extension of the region in which the lowest-energy charge excitation in the insulating phase is bosonic. The observed weak non-monotonicity of the crossover curve as the interaction strength decreases reflects the bosonic character of the lowest-energy charge excitations close to the SIT.

\section{Summary}\label{sec:summary}

We used unbiased numerical calculations to study the SIT in the ground state of the attractive honeycomb Hubbard model in the presence of a staggered potential. We directly showed that the lowest-energy charge excitations are bosonic across the transition, and cross over to fermionic in the insulating phase. The former is consistent with the finding that the SIT in this model belongs to the 3$d$-XY universality class. The results obtained in the honeycomb lattice are qualitatively similar to those in the square lattice~\cite{Mondaini2015}, suggesting that our findings are universal for SITs in clean systems. For example, we expect similar results for the attractive Hubbard model in the triangular lattice if one adds a positive (negative) local potential to one site in each triangle when the filling is $n=4/3$ ($n=2/3$).

Given that, in the absence of nearest-neighbor hoppings, the Hamiltonian considered here has already been simulated in optical lattice experiments~\cite{Messer2015}, we expect that our findings can readily be tested in such experiments. An interesting open question for both theory and experiments is what happens in three dimensions, in which the critical temperature that triggers the onset of superconductivity is nonzero at half-filling~\cite{Staudt2000, Sewer2002}. Exploring the interplay between the temperature and the parameters considered here in two dimensions may help improve our understanding of the role of the preformation of pairs in the finite-temperature realm.

\begin{acknowledgments}
Y.L.~was supported by the China Postdoctoral Science Foundation under Grants No. 2019M660432 and No. 2020T130046 as well as the National Natural Science Foundation of China (NSFC) under Grants No. 11947232. R.M.~acknowledges support from NSFC Grants No.~U1930402, 12111530010, 11974039, and 12222401; M.R.~acknowledges support from the National Science Foundation Grant No.~2012145. The computations were performed on the Tianhe-2JK at the Beijing Computational Science Research Center.
\end{acknowledgments}

\appendix

\section{Phase diagram in a 16-sites lattice}
\label{sec:16B}

\begin{figure}[!t] %%% FIG Appendix A
\includegraphics[width=0.9\columnwidth]{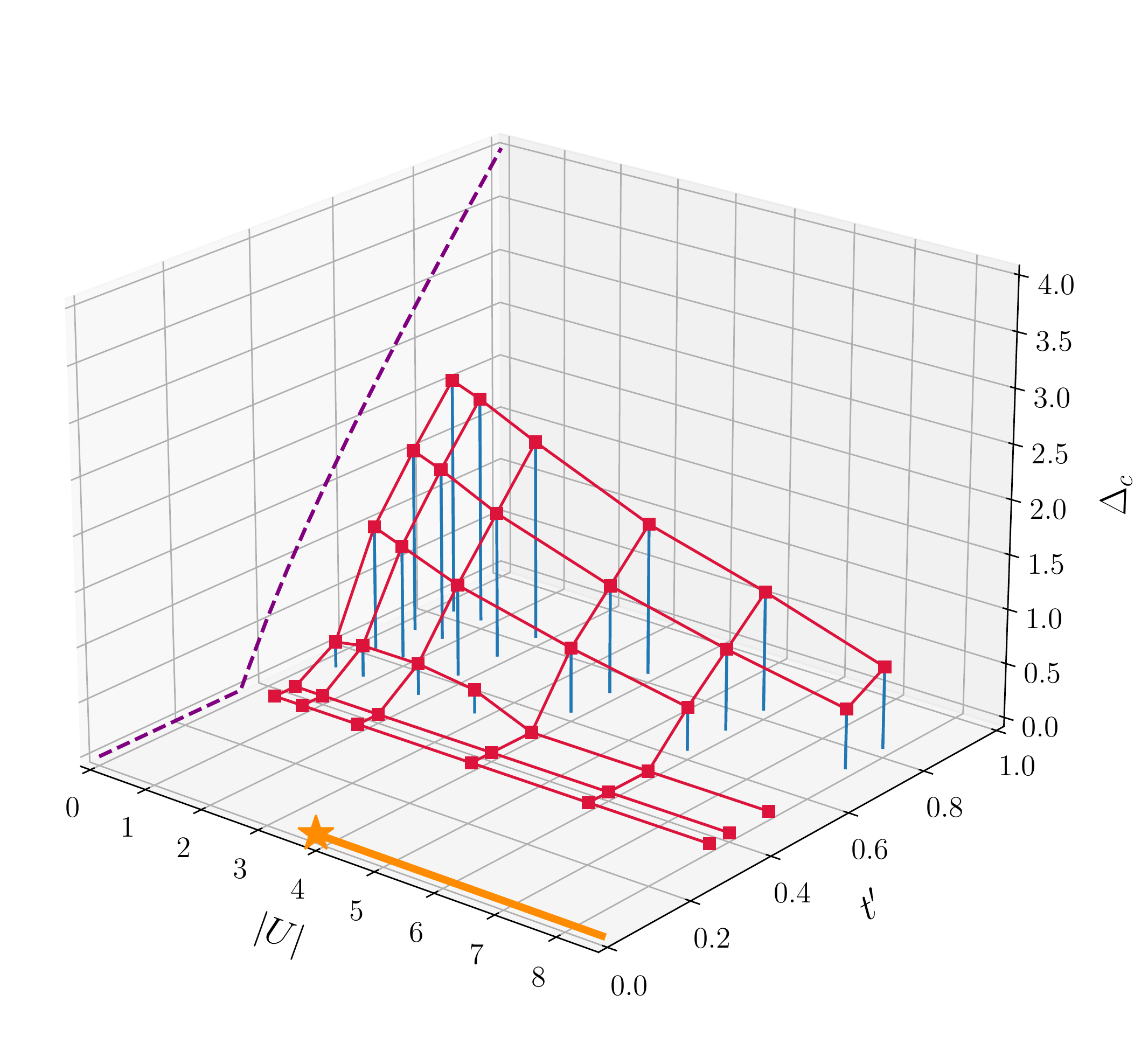}
  \vspace{-0.3cm}
  \caption{Phase diagram of the SIT transition as obtained using exact diagonalization in a 16-site lattice (lattice 16B in Ref.~\cite{Shao2021}).}
  \label{fig:ED_16B}
\end{figure}

In Fig.~\ref{fig:ED_16B}, we show the phase diagram obtained as explained in Sec.~\ref{sec:ed} using the fidelity susceptibility but in a lattice with 16 sites. This lattice geometry was referred to as cluster 16B in Ref.~\cite{Shao2021}, and does not feature the $K$ and $K^\prime$ high symmetry points of the Brillouin zone. Nonetheless, the resulting phase diagram closely follows the one reported in Fig.~\ref{fig:phase_diagram}(a), which was obtained in the commensurate 18-sites lattice depicted in Fig.~\ref{fig:cartoon_bands_dos}.

\section{Kinetic energy in PQMC} \label{sec:kin}

\begin{figure}[!t] %%% FIG Appendix B
  \includegraphics[width=0.9\columnwidth]{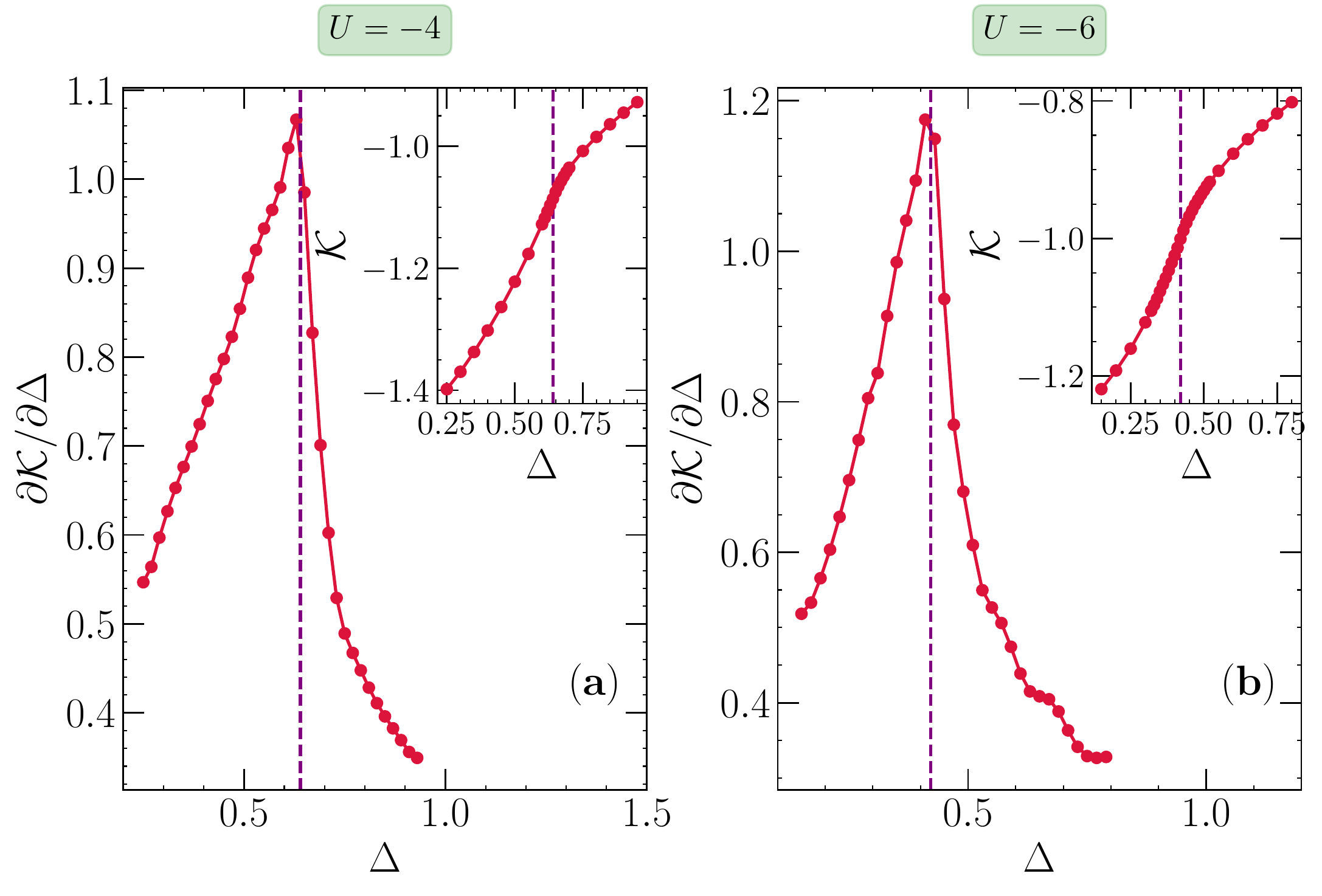}
  \vspace{-0.3cm}
  \caption{Numerical derivative of the kinetic energy per site $\cal K$ with respect to $\Delta$ in a lattice with $L=15$, for $t^\prime = 0.6$, when (a) $U=-4$ and (b) $U = -6$. Insets: $\cal K$ before differentiation. Vertical dashed lines mark the SIT location, $\Delta_c$, obtained using the scaling analysis of the pair structure factor.}
  \label{fig:kin}
\end{figure}

\begin{figure}[!b] %%% FIG Appendix C
% \vskip 0.2in
  \includegraphics[width=0.98\columnwidth]{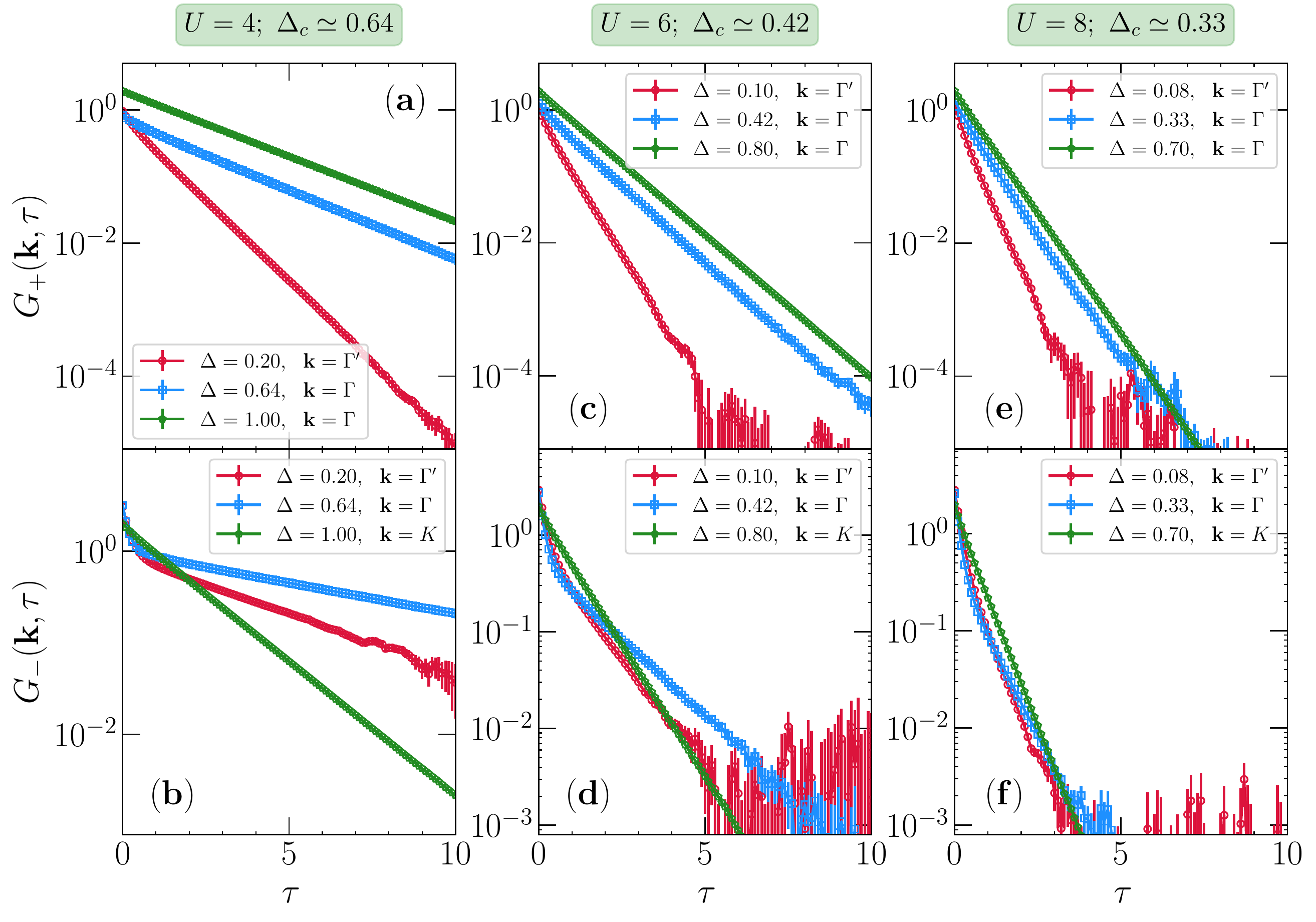}
  \vspace{-0.3cm}
  \caption{Dependence of the imaginary-time displaced Green's functions related to particle (upper row) and hole (bottom panels) excitations at $t^\prime = 0.6$ and increasing interaction strengths, $U = -4, -6$, and $-8$. For each interaction strength, we plot $G_\pm({\bf k}, \tau)$ for values $\Delta < \Delta_c$, $\Delta \simeq \Delta_c$, and $\Delta > \Delta_c$, for the momentum ${\bf k}$ corresponding to the smallest gap. In the superconducting regime, we observe a direct gap at the closest ${\bf k}$ point to the center of the Brillouin zone $\Gamma^\prime \equiv (0, 2\sqrt{3}\pi/9)$ for this system size, while at the transition point the direct gap resides at $\Gamma = (0,0)$. An indirect gap ensues within the insulating regime (see text). These results were obtained in a lattice with $L=6$.}
  \label{fig:G_decay}
\end{figure}

In Fig.~\ref{fig:PQMC_local_qtts}, we showed that the double occupancy $D$ (resolved in each sublattice) can be used as a proxy to locate the SIT in experiments. Other local quantities, such as the kinetic energy, can equally be used to locate the SIT in experiments. Figure~\ref{fig:kin} shows the variation of the kinetic energy per site ${\cal K}$ with increasing the strength $\Delta$ of the staggered potential, for two values of $U$, as obtained using PQMC simulations in a lattice with $L=15$. A prominent kink is observed in the derivative of ${\cal K}$ at $\Delta_c$.

\section{Single-particle gap in PQMC}\label{sec:G_tau}

In Sec.~\ref{sec:excitations}, we explained the procedure used to extract the one-particle gap based on the exponential decay of the one-particle Green's functions at long imaginary-times. In Fig.~\ref{fig:G_decay}, we show $G_\pm({\bf k}, \tau)$ for staggered potentials $\Delta < \Delta_c$, $\Delta \simeq \Delta_c$, and $\Delta > \Delta_c$, for each value of $U$, for the ${\bf k}$-points that give the smallest gap $\delta_\pm({\bf k})$ after fitting $G_\pm({\bf k}, \tau)$ to an exponential form at large $\tau$'s. Irrespective of the interaction strength, an overall trend can be seen in which a direct one-particle gap within the superconducting regime or at the transition point is replaced by an indirect gap, $K,K^\prime\leftrightarrow\Gamma$, within the insulating phase. As pointed out in the main text, in the noninteracting regime such an indirect gap is also observed for values of $\Delta > \Delta_c$ between precisely the same high-symmetry points.

\bibliography{attractive_IHM_honeycomb}

\end{document}